# Temperature-driven polarization rotation and triclinic phase at morphotropic phase boundary of Pb(Mg$_{1/3}$Nb$_{2/3}$)O$_3$ – PbTiO$_3$ crystals

Alexei A. Bokov*, Haiyan Guo and Zuo-Guang Ye†

*Department of Chemistry and 4D LABS, Simon Fraser University, Burnaby, BC, Canada*

Information about the crystal structures in the range of morphotropic phase boundary of ferroelectric perovskite solid solutions is important for understanding their intricate properties which result in wide opportunities for practical applications. However, for the (1- $x$)Pb(Mg$_{1/3}$Nb$_{2/3}$)O$_3$ – $x$PbTiO$_3$ solid solution system this information is contradictory. Different composition-temperature phase diagrams have been reported for this system in literature based on the investigations of single crystals and ceramics using various experimental techniques. In this work we apply polarized light microscopy (PLM), X-ray diffraction (XRD) and dielectric spectroscopy to study the crystal structure and phase transitions in the 0.68Pb(Mg$_{1/3}$Nb$_{2/3}$)O$_3$ – 0.32PbTiO$_3$ single crystal. We confirm the monoclinic M$_B$ symmetry (space group $Cm$) of the room-temperature phase. According to PLM, it transforms with increasing temperature into a triclinic (Tr) phase rather than the previously reported monoclinic M$_C$ or tetragonal phase. XRD data are consistent with the presence of Tr phase. The Tr phase transforms to the monoclinic M$_C$ ($Pm$) phase and then to the cubic phase. Ergodic relaxor behavior is observed above the Curie temperature. The unit cell in the M$_C$ phase is pseudotetragonal with the lattice parameters $a = b < c$ and small monoclinic angle $β$. In the M$_B$ phase the direction of spontaneous polarization is temperature independent and close to the <111> pseudocubic direction. In the Tr and M$_C$ phases it changes with temperature so that near the Curie point it is close to [001] axis. No significant anomalies in the dielectric properties or changes in the domain structure are observed at the M$_B$ → Tr and Tr → M$_C$ phase transitions. The domain structure changes dramatically when the temperature varies within the Tr phase, causing a sharp change in birefringence.

## I. INTRODUCTION

Solid solutions of lead oxide perovskites with morphotropic phase boundary (MPB), such as (1 – $x$)PbZrO$_3$ – $x$PbTiO$_3$ (PZT) and (1 – $x$)Pb(Mg$_{1/3}$Nb$_{2/3}$)O$_3$ – $x$PbTiO$_3$, are widely used ferroelectric materials [1-3]. MPB is a boundary, $x_B$, separating the compositions with different crystallographic symmetry, typically rhombohedral (R) and tetragonal (T). The value of $x_B$ is nearly independent of temperature, so that the MPB line appears nearly vertical on the composition-temperature phase diagram. Many physical properties, e.g. piezoelectric coefficients and dielectric constant, are maximized in compositions near MPB, making them attractive for applications. The enhanced properties were initially thought to be related to the coexistence of R and T phases in the MPB range [4,5], but later investigations, including powder neutron diffraction and high-resolution synchrotron X-ray diffraction measurements [6] have shown a more complex picture of MPB, where a narrow composition range with phases of lower, i.e. monoclinic or orthorhombic, symmetry is sandwiched between the R and T phases. Identifying these intermediate phases using traditional diffraction techniques is challenging, because their unit cells are only slightly distorted as compared to those of R and T phases, and phase coexistence is thought to be possible. In addition, the local crystal symmetry can differ from the macroscopic one, and different results can be obtained when analyzing the structure at different length scales, i.e. macroscopically (with polarized light microscopy), mesoscopically (with Bragg diffraction or piezoresponse force microscopy) or locally (with pair distribution function analysis, Raman scattering, first principles-based simulations, etc.). According to various proposed structural models, the symmetry of MPB phases in perovskite solid solutions may increase [7] or decrease [8] with increasing scale. To facilitate further discussion, the inset in Fig. 1 schematically shows the directions of spontaneous polarization in phases of all symmetries that can be observed in ferroelectrics with a cubic $Pm\bar{3}m$ paraelectric phase. These directions are related to crystal symmetry and are used to classify ferroelectric phases.

The system (1 – $x$)Pb(Mg$_{1/3}$Nb$_{2/3}$)O$_3$ – $x$PbTiO$_3$ (PMN-$x$PT) is a solid solution between the normal ferroelectric PbTiO$_3$ (PT) and the canonical relaxor Pb(Mg$_{1/3}$Nb$_{2/3}$)O$_3$ (PMN), and the compositions with $x$ around and below MPB exhibit some properties characteristic of relaxor ferroelectrics. Fig. 1 shows the phase diagram of the PMN-$x$PT system, summarizing the results of recent structural studies. The dependence of Curie temperature ($T_C$) separating the cubic (C) phase with space group $Pm\bar{3}m$ and the ferroelectric phase on $x$ is practically linear. The ferroelectric phase observed in solid solutions with comparatively large $x$ is undoubtedly tetragonal $P4mm$ (the same as in PT). Most papers [9-14] report the following sequence of phases observed at room temperature with decreasing $x$: tetragonal $P4mm \rightarrow$ monoclinic $\rightarrow$ rhombohedral $R3m \rightarrow$ cubic $Pm\bar{3}m$. However, the exact values of concentrations separating these phases ($x_{Bi}$) are controversial, as one can see in Table 1. Furthermore, some authors found that the monoclinic region of phase diagram consists of two monoclinic phases, $Pm$ (conventionally denoted as $M_C$) and $Cm$ ($M_B$), while in the phase diagrams by Noheda et al. [9] and Slodczyk et al. [14] the $Cm$ phase is missing (i.e. $x_{B1} = x_{B2}$). A significant discrepancy concerns the position of the $R3m$ phase. Polarized light microscopy (PLM) [12] and X-ray powder diffraction (XRPD) investigations revealed this phase at $x < x_{B1} \approx 0.3$ [11,15], but in neutron powder diffraction (NPD) experiments the monoclinic $Cm$ phase of $M_B$ type was observed in all studied solutions with $0.1 \leq x \leq 0.3$ and the rhombohedral phase was not observed at all [13, 18].

*Contact author: abokov@sfu.ca

†Contact author: zye@sfu.ca

One more controversial part of the phase diagram is the high-temperature area of MPB just below the Curie temperature (highlighted in blue in Fig. 1). Different temperature-induced phase sequences were suggested for this area in the literature. In one version of phase diagram based on the laboratory [14,15] and synchrotron [9] X-ray powder diffraction and the X-ray and neutron diffraction in single crystals [17], the low-temperature (rhombohedral or monoclinic) phase transforms on heating into the $P4mm$ phase and then into the cubic $Pm\bar{3}m$ phase. In the PLM studies of crystals, [12] the T phase was not observed at any temperature when $x < x_{B3}$. The $Cm \rightarrow Pm \rightarrow Pm\bar{3}m$ phase sequence was revealed at $x_{B1} < x < x_{B2}$, while $Pm$ phase transformed directly to $Pm\bar{3}m$ at $x_{B2} < x < x_{B3}$. According to the third version based on neutron powder diffraction data [13] three phase transitions exist upon heating at $0.27 < x < x_{B2}$ with the $Cm$ ($M_B$) $\rightarrow Pm \rightarrow P4mm \rightarrow Pm\bar{3}m$ phase sequence, and at $x_{B2} < x < x_{B3}$ the sequence is $Pm \rightarrow P4mm \rightarrow Pm\bar{3}m$.

An unusual behavior named "anomalous skin effect" was observed in single crystals (but not in ceramics or powders) of PMN-$x$PT and of another relaxor-based system $(1–x)$Pb(Zn$_{1/3}$Nb$_{2/3}$)O$_3$ – $x$PbTiO$_3$ with relatively small $x$. The crystal structure was found to be different inside the crystal and in the near-surface layer ("skin"), whose thickness was estimated to be approximately 10 - 50 μm [18]. The conclusion about the existence of skin was made based on a comparison of structural data obtained on crystals of the same composition using radiation sources of different energies and, therefore, different penetration depths. In particular, neutron diffraction experiments with PMN-$x$PT indicated that at low temperatures the bulk of a single crystal remained metrically cubic (lattice parameters $a = b = c$ and $\alpha = \beta = \gamma = 90°$) for $x \leq 0.2$ and exhibited short-range, pseudo-rhombohedral polar distortions for $0.2 \leq x \leq \sim 0.27$ [18-20]. This internal metrically cubic phase was called "phase X". The true symmetry of phase X, which is determined not only by the shape of the unit cell (lattice parameters), but also by ion positions inside the unit cell, has not yet been determined. In contrast, the symmetry of ceramics and powders, which is expected to be the same as the crystal skin symmetry, was found to be long-range rhombohedral or monoclinic for $0.05 \leq x \leq x_{B2}$, both in Xray and neutron diffraction experiments [11,18,21]. The cutoff concentration, below which the skin effect exists in PMN-$x$PT,

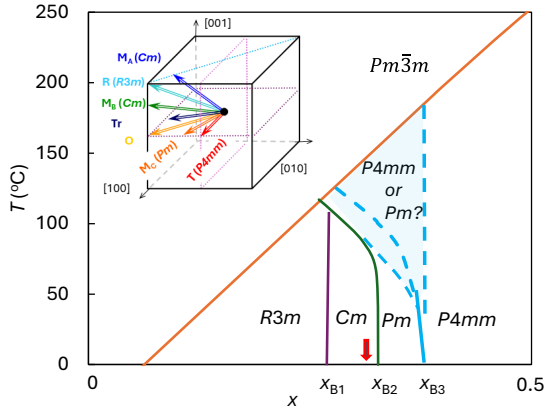

FIG.1 Composition-temperature phase diagram of the $(1 – x)$Pb(Mg$_{1/3}$Nb$_{2/3}$)O$_3$ – $x$PbTiO$_3$ solid solutions in the range of $x = 0 \sim 0.5$ based on Refs. [9-16,18]. The space group of each phase is indicated. The values of interphase boundary concentrations, $x_{Bi}$, reported by different authors may be different, as discussed in the text. The disputable position of the boundary between the monoclinic or rhombohedral phase and the tetragonal $P4mm$ phase is shown by dashed lines. Arrow indicates the composition $x = 0.32$ studied in this work. Inset shows the direction of the spontaneous polarization ($\boldsymbol{P}_S$) in the tetragonal (T), orthorhombic (O), rhombohedral (R), monoclinic (M$_A$, M$_B$ or M$_C$) and triclinic (Tr) phases with respect to the crystallographic axes of the paraelectric C phase. One of several symmetry-equivalent directions (orientation states) is shown for each phase. The crystal symmetry restricts $\boldsymbol{P}_S$ to be in the <100>, <110> and <111> directions in the T, O and R phases, respectively, and to lie in the {100} plane in the M$_C$ phase or the {110} plane in the M$_A$ and M$_B$ phases. The M$_A$ and M$_B$ phases are distinguished according to the $\boldsymbol{P}_S$ location in its {110} plane: $\boldsymbol{P}_S$ lies between <001> and <111> in the M$_A$ phase or between <111> and <110> in the M$_B$ phase. Any $\boldsymbol{P}_S$ direction is permissible in the Tr phase.

*Contact author: abokov@sfu.ca

†Contact author: zye@sfu.ca

TABLE I. Compositions of $(1 - x)$Pb(Mg$_{1/3}$Nb$_{2/3}$)O$_3$ – $x$PbTiO$_3$ solid solutions expressed as the concentration of PbTiO$_3$, $x_B$, at which the boundaries between the $R3m$ and $Cm$ ($x_{B1}$), $Cm$ and $Pm$. ($x_{B2}$), $Pm$ and $P4mm$ ($x_{B3}$) phases were observed at room temperature in ground ceramics or single crystals using X-ray powder diffraction (XRPD), synchrotron X-ray powder diffraction (SXRPD) and neutron powder diffraction (NPD), and in single crystals using polarized light microscopy (PLM).

| Reference | Material | Technique | $x_{B1}$ | $x_{B2}$ | $x_{B3}$ |
|---|---|---|---|---|---|
| Noheda *et al*. [9] | Ceram. | SXRPD | 0.305 | 0.305 | 0.345 |
| Singh & Pandey [10] | Ceram. | XRPD | 0.27 | 0.3 | 0.345 |
| Singh *et al*. [13] | Ceram. | NPD | < 0.2 | 0.305 | 0.35 |
| Phelan *et al*. [18] | Ceram. | NPD | < 0.1 | > 0.3 | < 0.4 |
| Slodczyk *et al*. [14] | Crystal | XRPD | ≈ 0.3 | ≈ 0.3 | ≈ 0.35 |
| Shuvaeva *et al*. [12] | Crystal | PLM | 0.295 | 0.36 | 0.47 |

was estimated to be $x_c \approx 0.31$ [18].

Explanation of the discrepancies found in the literature regarding the phases and phase transitions in the PMN-$x$PT solid solutions of nominally identical compositions is complicated not only by the difference in the experimental techniques used, but also by the closeness of free energies of different phases in the compositions close to MPB. As a result, the discrepancy in the observed phase symmetry may result from the difficulties in controlling the technical parameters of sample preparation. To overcome this uncertainty, we study in this work the structure and properties of physically the same PMN-xPT ($x$ = 0.32) crystal samples using the techniques of XRD and PLM, which probe the structure on different scales, and dielectric spectroscopy which helps to determine the phase transition temperatures. This composition is particularly interesting for investigations because three temperature-induced phase transitions can be expected therein, and it exhibits extraordinary piezoelectric [2] and energy storage [22] properties. We found the following sequence of phases upon heating: $M_B \rightarrow Tr \rightarrow M_C \rightarrow C$ and discussed the reasons why different sequences were reported in the literature.

## II. METHODS

Single crystals of $(1-x)$Pb(Mg$_{1/3}$Nb$_{2/3}$)O$_3$ – $x$PbTiO$_3$ with $x$ = 0.32 (PMN-32PT) were grown by the top-seeded solution growth method which was reported elsewhere [23]. The rectangular-shaped faces of as-grown crystals were parallel to the {001} crystallographic planes (all directions in this paper refer to the pseudocubic crystal axes). The specimens for measurements were cut parallel to the (001) natural faces to obtain platelets with the dimensions of ~ 2 × 2 × 0.5 mm. The large crystal faces were finely polished, and the crystals were annealed at 400 °C afterward to remove any residual stresse. Gold electrodes were sputtered onto large faces for dielectric and piezoelectric investigations.

Piezoelectric coefficient $d_{33}$ was measured by the quasistatic Berlincourt method using a ZJ-6B $d33/d31$ meter. The real part of relative permittivity ($\varepsilon$) and dielectric loss (tan $\delta$) were measured at a field strength of about 2 Vmm$^{-1}$ using a Novocontrol Alpha broadband dielectric analyzer equipped with a Quatro Cryosystem for temperature control. The data were collected upon heating or cooling the specimen at a rate of 0.5 K min$^{-1}$.

X-ray diffraction (XRD) investigations were carried out using a high-resolution Bruker D8 Advance diffractometer with filtered Cu K$\alpha$ radiation. The diffractometer was equipped with a silicon strip detector and a thermal stage for high-temperature measurements. The data were collected from the (100) surface of the crystal plate in Bragg-Brentano geometry. This type of scan allows to determine the interplanar $d_{h00}$ spacing from the h00 peak positions using Bragg's law, but an additional zero shift can appear if the diffraction vector of the diffractometer is not exactly perpendicular to (100) crystal plane. This can happen e.g. because the large faces of crystal plate are not exactly parallel to (100). To correct the zero shift, we added to the recorded $2\theta$ values a constant correction ($\Delta 2\theta$), which was determined so that the $d_{h00}$ values calculated at room temperature from the 300 and 400 reflections were the same. Typically, in different samples we found $|\Delta 2\theta| \sim 0.2°$ or smaller. To determine the reflection $2\theta$ positions accurately, we fitted the measured Bragg peak profile to a sum of pseudo-Voigt peaks and a linear background with the help of WinPLOTR program [24]. QualX program [25] was used to remove the K$\alpha_2$ component from the peak profiles.

*Contact author: abokov@sfu.ca

†Contact author: zye@sfu.ca

The polarized light microscopy (PLM) studies were performed using an Olympus BX60 polarizing microscope equipped with an Olympus UC 30 digital camera, the Berek U-CTB and U-CBE compensators and a first order red plate (530 nm). Optical birefringence was calculated according to the relation $\Delta n = \Gamma/t$, where $t$ is the thickness of the crystal plate and $\Gamma$ is the retardation measured by the compensator. This value of $\Delta n$ is an effective birefringence which may represent the intrinsic property of the material, but also depend on the domain structure. A Linkam HTMS600 heating/cooling stage was used for temperature-variable.

### III. RESULTS AND DISCUSSION

#### A. Dielectric measurements

The dielectric behavior of the studied single crystal is shown in Figure 2. The temperature dependence of relative permittivity, $\varepsilon$, is qualitatively similar to those previously published for the PMN-PT single crystals of the same composition [26,27]. The large $\varepsilon(T)$ peak observed at a temperature of about 150°C is associated with $T_C$. The composition of (1-$x$)PMN-$x$PT solid solution can be determined using empirical relations $x = (T_C + 10)°C/500°C$ [28] or $x = (T_C – 236)$ K/586 K [29]. Both confirm that $x = 0.32$ in our crystals. Three temperature ranges with different behavior can be distinguished below $T_C$. In the ranges from $T_C$ to approximately 110 °C and from room temperature to 80 °C (on heating) or 50 °C (on cooling), the permittivity and tan $\delta$ change slightly, but between these two ranges a significant change in dielectric properties is observed. Our structural studies reported below indicate that these three temperature ranges correspond to three ferroelectric phases of different symmetries.

In the vicinity of $T_C$, characteristic features of crystals undergoing a spontaneous relaxor-to-ferroelectric phase transition [30] are evident. At temperatures above the sharp change of permittivity, which points to a first order ferroelectric phase transition, a diffuse maximum is observed whose magnitude decreases and whose temperature, $T_m$, increases with increasing measurement frequency ($f$), suggesting a dielectric relaxation process. In relaxor ferroelectrics the dependence of $\varepsilon$ on $T$ in a wide temperature range above $T_m$ can be described by the empirical equation,

$$\frac{\varepsilon_A}{\varepsilon} = 1 + \frac{(T-T_A)^\gamma}{2\delta^\gamma}, \qquad (1)$$

where $T_A$, $\varepsilon_A = \varepsilon(T_A)$, $\delta$ and $\gamma$ are the adjustable parameters. The diffuseness parameter, $\delta$, characterizes the width of the permittivity peak. It has been concluded [31] that the parameter $\gamma$ equals two for all relaxors, regardless of the degree of diffuseness. Our results agree with this conclusion. We fit the $\varepsilon(T)$ dependence measured at 100 kH to Eq. (1) in the temperature range between ($T_m$ + 3) = 155 °C and 300 °C. The best-fit value $\gamma = 2.04 \pm 0.05$ is found. The fit with the fixed $\gamma = 2$ is demonstrated in Fig. 2(b). The best-fit values $T_A = 143.4 \pm 0.9$ °C and $\delta = 18.4 \pm 0.4$ K are found in this case.

The frequency dependence of $T_m$ in relaxors should follow the Vogel-Fulcher (VF) law:

$$f = f_0 \exp\left[E_a/(T_{VF} - T_m)\right], \qquad (2)$$

where $f_0$, $E_a$ and $T_{VF}$ are the adjustable parameters. Fig. 1(c) demonstrates the fulfillment of this law for the $T_m$ temperatures determined during cooling of the crystal. The best-fit parameters are determined to be $f_0 = 5\times10^{12}$ Hz, $E_a = 184 \pm 3$ K and $T_{VF} = 142.5 \pm 1.0$ °C. Upon heating the VF law cannot be reliably verified. In this case, at relatively low frequencies the position of permittivity maximum is determined by the phase transition (i. e. $T_m = T_C$), rather than by the relaxation process, and the dependence of $T_m$ on $f$ appears only at $f$ higher than 10 kHz.

The fulfillment of Eq. (2) may be associated with the glassy freezing of dipole dynamics at $T_{VF}$, but not necessarily [32, 33]. As shown in Ref. 32, this relation can be due to a gradual broadening of the relaxation spectrum, when the total relaxator strength of the system (static permittivity) reaches the maximum at the temperature $T_A \approx T_{VF}$. We cannot determine whether the observed VF behavior is caused by glassy freezing, since this requires studying the temperature dependence of the relaxation frequency at $T > T_C$. In PMN-32PT this frequency is higher than the highest measurement frequency achievable for our dielectric analyzer.

*Contact author: abokov@sfu.ca

†Contact author: zye@sfu.ca

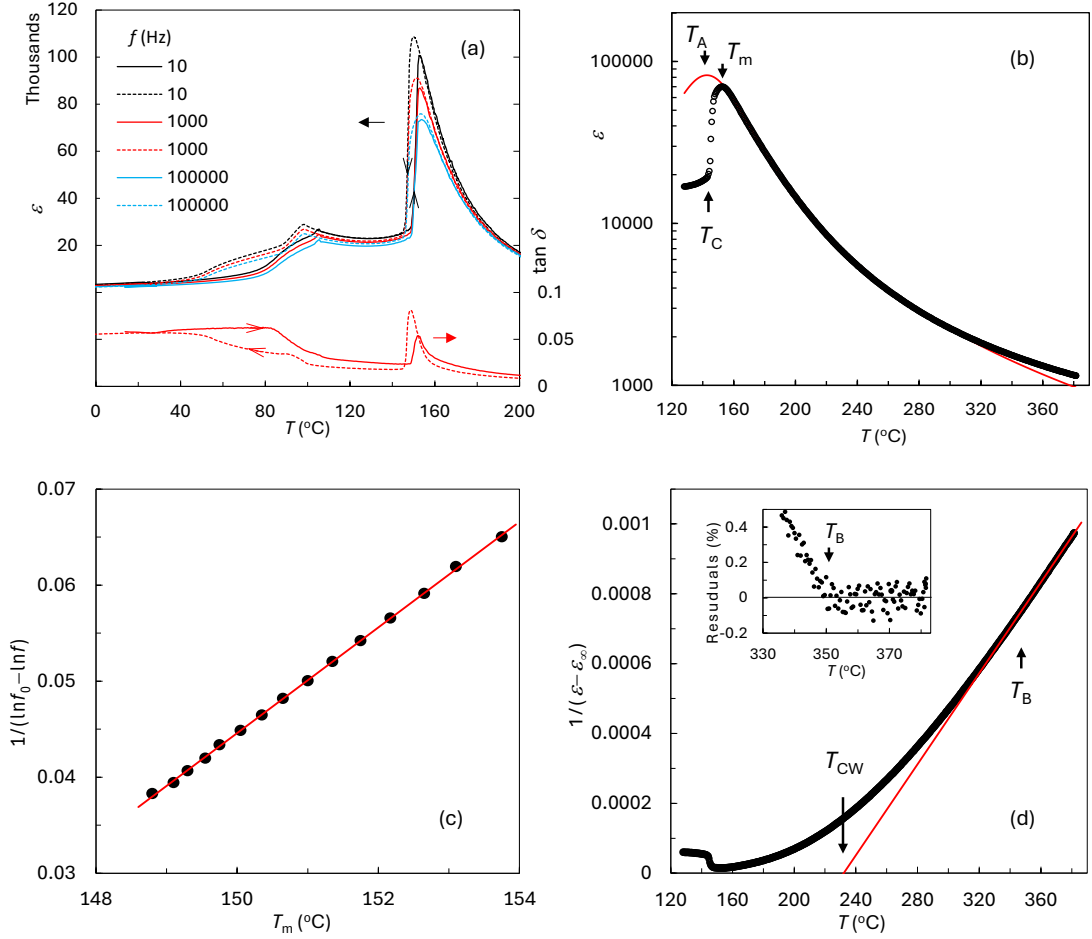

FIG. 2. Dielectric properties measured in (001) oriented PMN-32PT single crystal. (a) Temperature dependences of the relative permittivity and dielectric loss at different frequencies upon heating (solid lines) and cooling (dashed lines). (b) Fit of the relative permittivity measured at 100 kHz to Eq. 1 with $\gamma = 2$. (c) Fit of the permittivity maximum temperature measured at various frequencies between 10Hz and 460 kHz to Eq. 2. (d) Fit of the relative permittivity measured at 100 kHz to Eq. (3). The inset shows the fitting residuals found as $(\varepsilon_{cal} - \varepsilon)/\varepsilon$, where $\varepsilon_{cal}$ is the calculated value given by Eq. (3) with the best-fit parameters. In panels (b-d) the red solid line is the least-squares fit to the data points shown by circles.

At temperatures much higher than $T_m$ the permittivity in relaxors does not obey Eq. (1) and the Curie-Weiss (CW) law is typically observed [34, 35]:

$$\varepsilon = \varepsilon_\infty + C/(T - T_{CW}) \qquad (3).$$

Fig. 2 (d) illustrates the CW behavior in PMN-32PT crystal with the parameters $\varepsilon_\infty = 121$ and $C = 1.54 \times 10^5$ K typical of perovskite oxide relaxors. The value of the Curie-Weiss temperature $T_{CW} = 232$ °C is larger than $T_m$, which is also characteristic of relaxors. The insert in Fig. 2 (d) shows that the temperature $T_B = 350$ °C (known as the Burns temperature) limits the validity range of the CW law. The Burns temperature in relaxors separates the high-temperature paraelectric state and the low-temperature ergodic relaxor state [30]. Note that the value of $T_B$ which we found in PMN-32PT is practically the same as that in pure PMN and in PMN-35PT crystals [36]. This means that the $T_B$ in PMN-$x$PT solid solutions is independent of PT concentration. Similar behavior has been found in some other perovskite solid solutions [34].

It is widely believed that the relaxation dynamics in the ergodic relaxor phase is determined by its complex nanostructure, which includes dynamic nanoscale regions of spontaneous polarization, known as polar nanoregions (PNRs) [30, 36, 37]. The existence of dynamic PNRs at temperatures between

*Contact author: abokov@sfu.ca

†Contact author: zye@sfu.ca

$T_C$ and $T_B$ in the PMN-32PT crystals has been confirmed in Ref. 38 using neutron scattering experiments.

### B. X-ray diffraction

In our experiments we use the $\theta$-$2\theta$ scan to observe the diffraction from the set of (001) lattice planes parallel to the crystal surface. The distances between these planes can be different in different ferroelastic domains, being dependent on the domain orientations. Accordingly, for differently oriented domains Bragg's law is satisfied at different angles $\theta$, which results in the splitting of $h$00 Bragg reflections observed in a multidomain crystal. These observations are not sufficient to unambiguously determine the phase symmetry but may restrict the set of possible crystal systems. In particular, a single $h$00 peak observed in the parent cubic phase is expected to remain single in the R phase, split into two peaks in the T, O, $M_A$ and $M_B$ phases and into three peaks in the $M_C$ and Tr phases.

Figure 3 shows the result of the XRD measurements of PMN-32PT crystal. Only one family of $h$00 Bragg peaks is observed in the diffraction pattern (the peaks in this paper are denoted with pseudocubic indexes), which confirms that the surfaces of the crystal plate are parallel to the (100) planes. To determine the crystal symmetry at different temperatures, we analyze the higher-angle 400 peak, for which the peak splitting in the low symmetry phases should be most pronounced. We collected XPD data at several temperatures upon cooling and fitted the profiles to a sum of pseudo-Voigt shape functions. The fitting results are shown for selected temperatures in Fig. 3 and in more detail in Fig. S1 of Supplemental Material [39].

At the temperature of 300 °C a resolution-limited 400 singlet is observed [Fig. 3(f)], which can be well fitted to a Lorentzian function with a Full Width at Half Maximum (FWHM) of 0.083°. This is expected for a high-quality crystal of the cubic symmetry. However, at all temperatures below the Curie point the peak is clearly split (Fig. 3b-d).

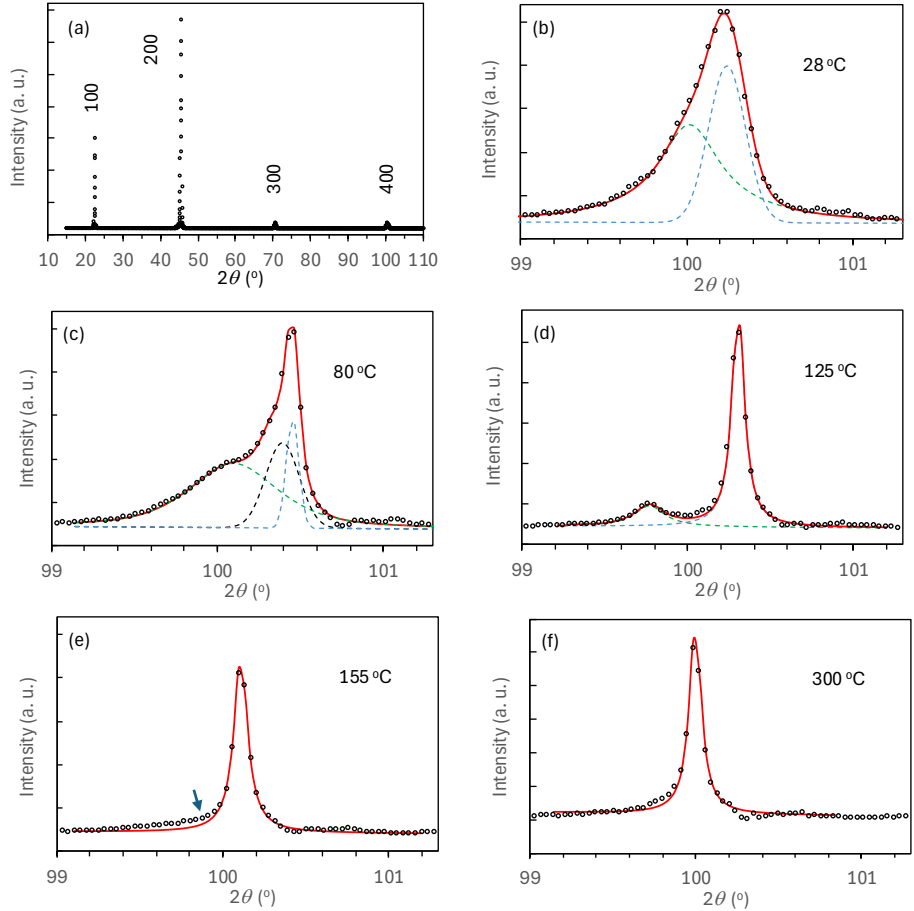

FIG. 3. Selected XRD patterns obtained from a main facet of the (001) oriented PMN-32PT crystal plate during slow cooling and indexed with respect to pseudocubic perovskite axes. (a) Full pattern at room temperature. (b-f) Profiles of the 400 Bragg peak in Phase IV at 28 °C (b), Phase III at 80 °C (c), Phase II at 125 °C (d), Phase I at 155 °C, (e) and Phase I at 300 °C (f). The K$\alpha_2$ component is removed from peak profiles using a standard software. The red solid line is the least-squares fit to the data points shown by circles, and the dashed lines are fitted pseudo-Voigt subpeaks.

*Contact author: abokov@sfu.ca  
†Contact author: zye@sfu.ca

The interplanar spacings were calculated from the pseudo-Voigt peak positions. Since in perovskite oxides the deviation of the primitive unit cell from the rectangular cuboid shape is known to be very small, the interplanar $d_{100}$, $d_{010}$ and $d_{001}$ spacings are practically equal to the parameters of the primitive perovskite cell, $a$, $b$ and $c$, respectively.

The dependences of the best-fit parameters on temperature are shown in Fig. 4. It should be underlined that the $\theta$-$2\theta$ scan which we use in our experiments consists of scanning within a single plane, while the Bragg diffraction is an intrinsically three-dimensional phenomenon. The peak positions are determined by the coordinates of the reciprocal lattice nodes in reciprocal space, which may be slightly away from the scanning plane. This may lead to significant errors in the observed positions, intensities and widths of the peaks. To obtain more accurate XRD data, three-dimensional reciprocal space mapping of the peak splitting should be performed, which is not possible with our diffractometer.

Four temperature intervals with different behavior can be distinguished in Fig. 4 and associated with different phases. For the sake of convenience, we denote these phases by Roman numerals as shown in Fig. 4.

In the low-temperature Phase IV, the peak is highly asymmetric and can be modeled with two pseudo-Voigt components [Fig. 3 (b)]. This excludes the rhombohedral and monoclinic $Pm$ ($M_C$) symmetries, in which a singlet and a triplet should be observed, respectively, but is consistent with $Cm$ ($M_A$ or $M_B$). The monoclinic $M_B$ phase has been found using XRPD [10] and NPD [13] at room temperate in the PMN-xPT ceramics with a slightly lower concentration $x = 0.30$ and the lattice parameters $b = 4.0123$ Å, and $a = c = 4.0223$ Å [10]. The parameters $b = 4.0155$ Å, and $a = c = 4.0230$ Å which we derived for our PMN-32PTcrystals are in reasonable agreement with those values as well as with the XRD values of $b = 4.0172$ Å, and $a = c = 4.0227$ Å reported for the PMN-30PT ceramics in Ref. [31]. Previous PLM investigations of PMN-32PT have revealed the $Cm$ symmetry [12], in agreement with our results. We conclude, therefore, that Phase IV in our crystals is monoclinic $Cm$, but we cannot determine, based on our XRD data, whether it is $M_A$ or $M_B$.

Figure 4 (a,b) shows that with increasing temperature up to about 60 °C, the parameters of the perovskite cell and, consequently, the cell volume, decrease, but then change sharply to an increasing trend. In addition, two peaks cannot model the 400 profile anymore; three peaks are needed [Fig. 3 (c)]. This behavior clearly points to a transition from $Cm$ to monoclinic $M_C$ or triclinic Phase III. The $Cm \rightarrow M_C(Pm)$ transition has been reported previously in the MPB range of PMN-PT [Fig. 1(a)] both for single crystals [12] (based on PLM) and for ceramics [13] (based on diffraction data). However, the reports on further phase evolution with increasing temperature are contradictory. The diffraction data [13] in ceramics revealed an intermediate T phase located between *Pm* and C phase, but according to the PLM data, the *Pm* phase in single crystals on heating directly to the C phase [12]. Our XRD results suggest the existence of an intermediate phase in the single crystals (which we denote as Phase II). The 400 profile in this phase consists of two well separated symmetric and comparatively narrow peaks [Fig. 3(d)], as expected for the T phase. The attempts to fit the pattern with three peaks led to the disappearance of one of the peaks. The temperature dependences of the lattice parameters in Phase II practically coincide with those found in the ceramics of the same composition [13], as shown Fig. 4(a).

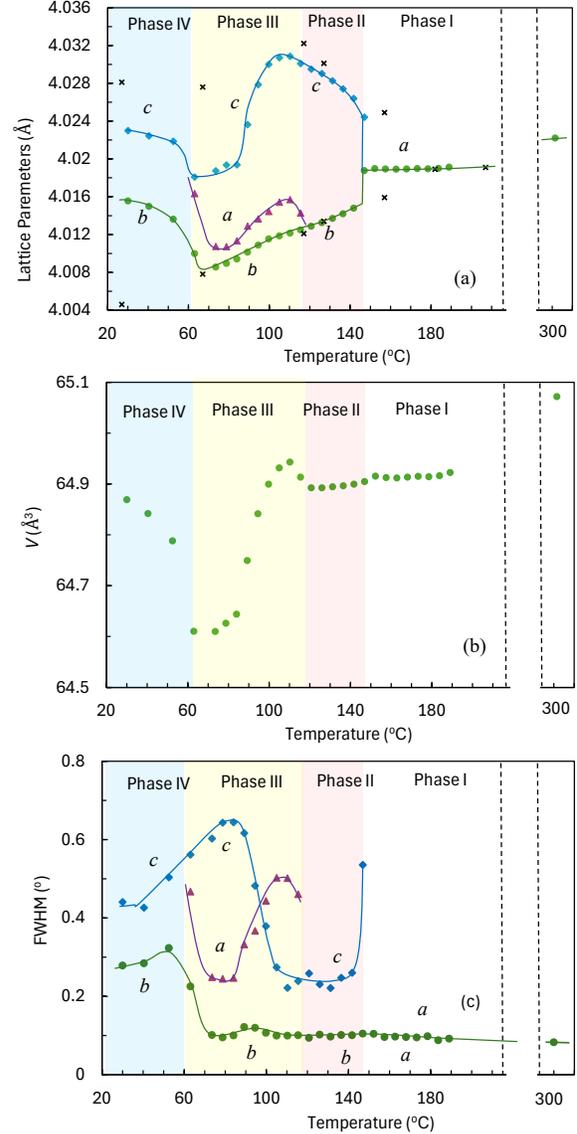

FIG. 4 Temperature dependences of the structural parameters obtained from XRD analysis of the PMN-32PT crystal. (a) Primitive perovskite cell parameters. The parameters for the PMN-32PT ceramics derived from powder newton diffraction data in Ref. [13] are shown by crosses for comparison. (b) Primitive perovskite cell volume. (c) Full Width at Half Maximum of the subpeaks derived from fitting the 400 Bragg peak. The subpeaks are labeled according to the corresponding lattice parameters. The temperature ranges of various phases that appear during cooling, obtained from the XRD data, are indicated.

*Contact author: abokov@sfu.ca

†Contact author: zye@sfu.ca

Note that in the PMN-33PT crystal, the symmetry of Phase II was found to be T if it was cooled from the C phase, but $M_C$ if it was heated from the lower-temperature phase [41]. To exclude similar behavior in our PMN-32PT crystals, we performed additional XRD measurements at several temperatures in Phase II obtained by heating the crystal from Phase III. The results were the same as for Phase II obtained from the cooling process: the 400 reflection remained a doublet. Fig. S2 illustrates this fact for one of the temperatures.

Above $T_C$, the crystal is in cubic Phase I, as evidenced by the single 400 peak [Fig. 3 (e)], which is only slightly wider than that observed at high temperatures (at 155 °C FWHM = 0.096). However, a small shoulder is observed at the low-angle base of the peak [marked by the arrow in Fig. 3(e)]. The intensity of the shoulder gradually decreases with increasing temperature, but it still remains at temperatures as high as 300 °C [Fig. 3(f)]. Similar shoulder at 300 °C near the 200 Bragg reflection was previously observed in the PMN-32PT crystals and attributed to the diffuse scattering caused by PNRs existing within the macroscopically cubic phase [14]. This explanation is consistent with the results of our dielectric measurements which suggest the presence of the ergodic relaxor phase at $T > T_C$. PNRs are expected in this phase. Interestingly, the authors of Ref. [14] reported that significant diffuse scattering tails were retained in low-temperature ferroelectric phases and interpreted this fact as evidence for the PNRs with different symmetries. However, we did not observe diffuse scattering below $T_C$ in our crystals [Fig 3 (b-d)].

In spite of significant and temperature dependent distortion of the perovskite cell in Phase II, the cell volume remains practically unchanged across this phase and over a wide temperature range above $T_C$ [Fig. 4(b)]. Such a kind of temperature-independent behavior (known as Invar effect) is characteristic of relaxor crystals [18].

The temperature hysteresis of phase transitions, observed in the dielectric measurements and confirmed by the PLM investigations presented in the next section, indicates that the transitions are of first order. A temperature range in which two phases coexist can be expected around first order transitions. The phase coexistence was indeed reported in the previous works [9, 10, 13]. It should result in the appearance of additional peaks in the XRD pattern, which could be resolved if they were sufficiently intense and separated. In an attempt to reveal possible phase coexistence we fitted the 400 profile to four peaks, but did not obtain meaningful results. However, we obtained indirect evidence. We recorded the goodness of fit parameter, $\chi^2$, calculated by the fitting software, as a function of temperature (see Fig. S3 in Ref. 39) and observed significant maxima near Phase I to Phase II and Phase III to Phase IV transitions. These maxima indicate that the fits in the transition regions are not entirely adequate, most probably due to the phase coexistence. No anomalies are observed near the transition from Phase II to Phase III, apparently because the lattice parameters in these phases are close to each other.

### C. Polarized Light Microscopy

The methods of optical crystallography are very helpful in studying the symmetry of ferroelectric phases, where deviations of the lattice parameters from the parameters of higher-symmetry phases can be very small and hardly detectable by diffraction methods. In particular, polarized light microscopy allows to find the directions of principal axes of optical indicatrix ellipsoid which are related to the crystal symmetry and spontaneous polarization direction. Because of symmetry requirements in the T, R and O phases, $P_S$ and indicatrix axis are parallel, while two indicatrix axes and $P_S$ lie in the same {110} plane in the $M_A$ or $M_B$ phases and in the same {100} plane in the $M_C$ phase. On the other hand, the Tr phase does not impose any restrictions on $P_S$ and indicatrix axes directions. In the PMN-PT crystals, which are proper ferroelectrics, the lattice deformation and the anisotropy of properties are caused by $P_S$, so we can assume that in all phases $P_S$ and one of the indicatrix axes are parallel to one another, at least approximately. This assumption should not affect the qualitative validity of our conclusions.

*Contact author: abokov@sfu.ca

†Contact author: zye@sfu.ca

We determine the direction of indicatrix axes, and, consequently, the phase symmetry by observing the angles $\delta$ between the crystallographic pseudocubic axis (axis of the parent cubic phase) and the crossed polarizers at which the crystal is in extinction. To facilitate further discussion, Fig. 5 demonstrates the extinction positions allowed in a (001) crystal platelet of different symmetries.

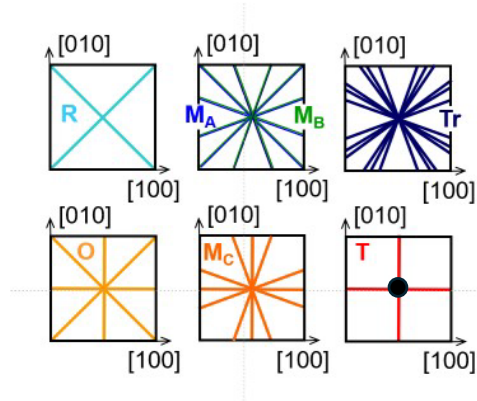

FIG. 5. Schematics of the positions of extinctions that can be observed under crossed polarizers (colored lines) in the different phases on a (001)-oriented crystal plate cooled from the cubic paraelectric phase. The black dot for T phase means that in some domains the extinction is observed at any position of crossed polarizers. The directions of pseudocubic crystallographic axes are indicated.

An additional indication for the phase symmetry is the position of ferroelectric/ferroelastic domain walls (angles $\varphi$ between the wall and crystallographic axis of the parent cubic phase). The permissible angles are determined by the mechanical compatibility conditions and the requirement for wall electrical neutrality. Significant deviations from the permissible wall positions would lead to an increase in elastic and electrostatic energies and thereby are usually not observed in ferroelectric phases. The list of domain wall directions for the crystals with cubic paraelectric phase can be found e.g. in Refs. [42, 43].

Examples of the domain structure observed in the different phases of PMN-32PT crystal are shown in Fig. 6. In Phase I almost the entire crystal remains in extinction (black) at any position of polarizers as Fig. 6(a) and Fig. S4 of Supplemental Material [39] demonstrate. However, in a relatively thin (~ 50 μm) layer near the lateral facets of the crystal plate, the extinction is observed only at $\delta = 0/90°$, which indicates that this layer is birefringent and, therefore, has a non-cubic symmetry different from the crystal bulk. This behavior is reminiscent of the skin effect found in PMN-$x$PT single crystals with relatively small $x$ (see Introduction). The birefringence of this near-surface layer ("skin") is estimated to be $\Delta n$ ~ 0.0002 and does not noticeably depend on temperature (see Fig. S4 of Supplemental Material for more details). Note, however, that this effect is sample dependent: in some other PMN-32PT crystals we did not observe the skin.

*Contact author: abokov@sfu.ca

†Contact author: zye@sfu.ca

When the temperature changes through the Curie point, the phase transition at different parts of the crystal occurs at slightly different temperatures. Fig. 6 (b) demonstrates this behavior, where the central (black) part of the crystal is in the cubic Phase I, while in the peripheral regions the transition to Phase II has already happened. At temperature 1 K lower, practically the entire crystal is transformed into phase II [Fig. 6 (c)]. This behavior is due to the well-known effect of composition gradient often observed in perovskite solid solution crystals and in PMN-PT in particular [12,44]. The titanium concentration, $x$, in the parts that crystallize later during the crystal growth process, appears to be greater, and, therefore, the Curie temperature in the near-surface layers of an as-grown crystal is higher than in the bulk. According to the $T$-$x$ phase diagram of PMN-PT, the $T_C$ difference of ~1 K which we observe corresponds to the $x$ difference of about 0.001. This relatively small composition inhomogeneity results in the sharp $\varepsilon(T)$ peak observed at $T_C$ (Fig. 2), whose maximum value of ~ $9 \times 10^4$ is significantly larger than the values of ~ $5 \times 10^4$ reported for the PMN-32PT crystals in some other papers.

Taking into account the composition gradient, we can also explain the above-discussed existence of birefringent near-surface layer (skin) in Phase I. In this layer the value of $x$ is larger than inside the crystal and, therefore, the lattice parameter is smaller. The mismatch of the lattice parameters results in lateral internal stress/strain and effective tetragonal symmetry of the skin. On the other hand, our as-grown crystals were thinned and then polished before observations, and the similar skin near large faces of the crystal platelet (perpendicular to the observation direction), even if it existed, was removed.

In accordance with the results of dielectric measurements (Fig. 2) we observed optically the $T_C$ hysteresis of about 4 K ($T_C$ = 151 – 152°C on heating and 147 – 148°C on cooling), indicating the nature of a first order phase transition. Another effect related to the first order character of the transition is demonstrated in Fig. S5 [39], which shows the development of the ferroelectric phase at a constant temperature slightly below $T_C$. Since the free energies of para- and ferroelectric phases around $T_C$ are close, but separated by a small energy barrier, significant time (more than 10 minutes) is needed for the phase transformation to be completed under isothermal conditions.

Fig. 6 (c) shows the crystal at $\delta = 45°$ just below the Curie temperature, where no domain walls are observed. The regions of different colors (including black) are not ferroelectric domains. These are interference colors due to different optical retardations related to the overlap of multiple submicron ferroelectric domains that are difficult to distinguish with an optical microscope. This kind of domain structure is often observed in ferroelectric phases of relaxors [45]. In the temperature range of approximately 2 K below $T_C$ a significant change in the domain structure is observed, and visible domain walls appear in several regions of the crystal, but upon further cooling the domain structure of Phase II changes slightly. This observation is consistent with the XRD data which indicate [Fig. 4(c)] a large FWHM of c domains around $T_C$ (suggesting their small thickness) and a sharp decrease in FWHM upon cooling (suggesting an increase in thickness).

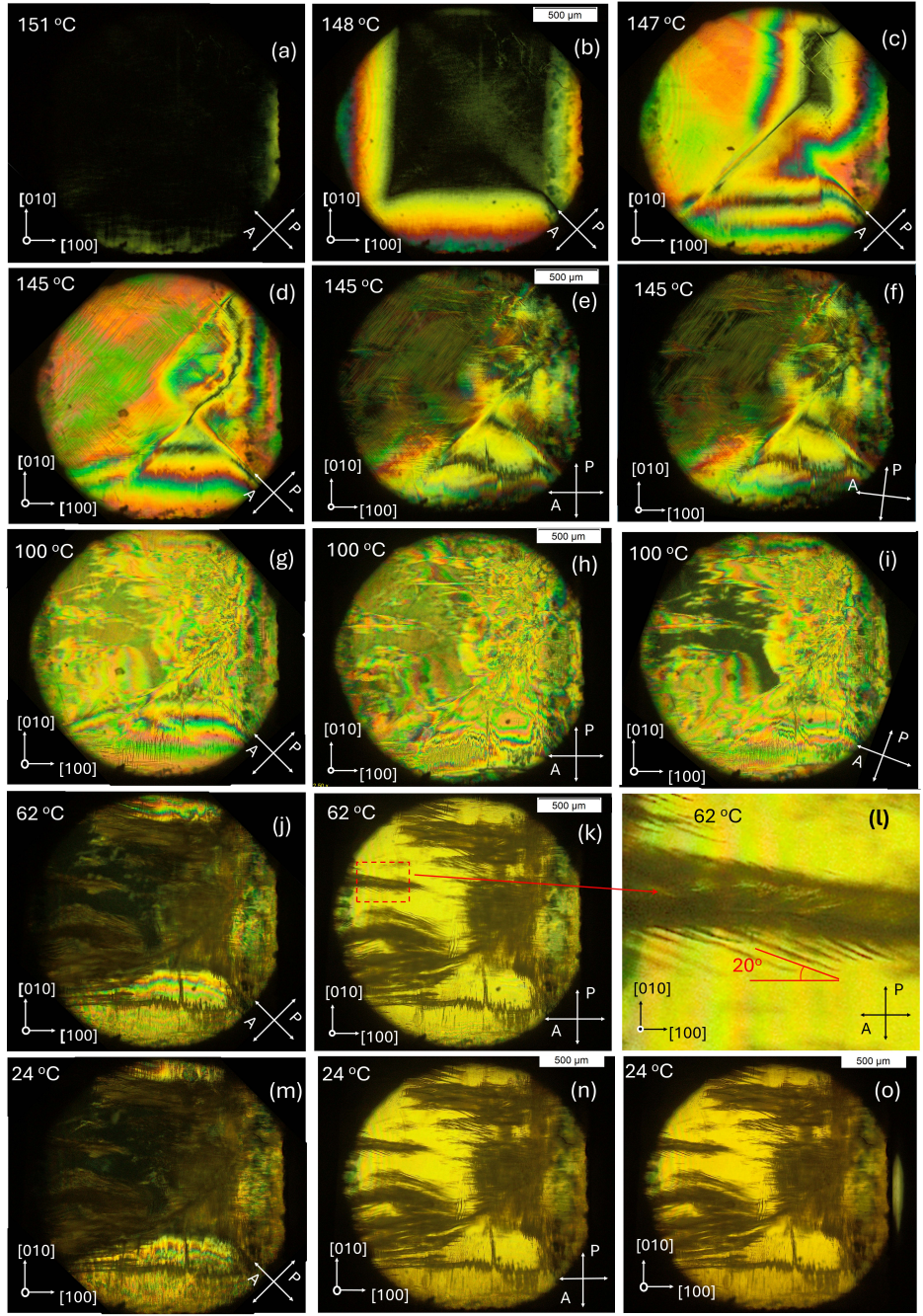

FIG. 6. Polarized light microscopy images of the same (001)-oriented PMN-32PT crystal platelet as was studied with XRD (in Figs. 3 and 4), obtained at different temperatures during cooling from 370 °C in Phases I (a-b), II (c – f), III (g -i) and IV (j – o). The temperature, directions of pseudocubic crystallographic axes and crossed polarizers are indicated. The platelet thickness is 540 μm.

*Contact author: abokov@sfu.ca

†Contact author: zye@sfu.ca

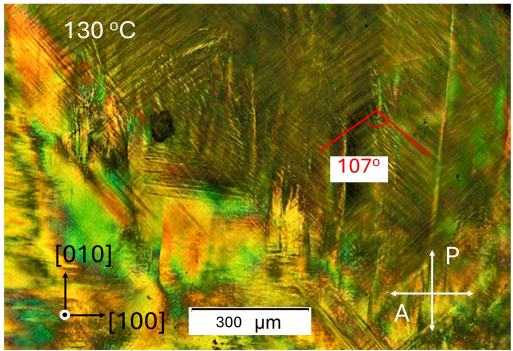

FIG. 7. Domain structure in Phase II of the (001)-oriented PMN-32PT crystal plate. Red lines parallel to some domain walls and the angle of 107º between them impermissible in the T phase are shown. The temperature, directions of pseudocubic crystallographic axes and crossed polarizers are also indicated.

Fig. 6 (d – f) shows the crystal in Phase II with different positions of crossed polarizers. Certain regions exhibit no extinction (i.e remain bright) at any δ. This behavior is incompatible with the R and T symmetries, at which all domains must be in extinction at δ = 45º and δ = 0º/90º, respectively (Fig. 5). Regions without extinctions may appear only in lower-symmetry phases due to the overlap of two or more domains in the path of the light ray. If the extinction angles in the overlapping domains are different, no extinction is observed at all. We can also see that in some regions of Phase II the extinction is observed at δ = 0/90º [Fig. 6 (e)], while in some others it occurs at δ = 7º. [Fig. 6 (f)]. According to Fig. 5, coexistence of domains with these two extinction angles is possible only in a monoclinic $M_C$ phase.

Another proof for the monoclinic symmetry of Phase II comes from the analysis of domain wall directions. Fig. 7 shows the domain structure of this phase in the same crystal obtained by heating from room temperature. The domains are larger in this case, and the domain walls are more visible. In the T phase the angle $\varphi$ between the walls traces on the (001) crystal surface and the crystallographic axes should be 45º or 0/90º. However, in Fig. 7, besides the walls with $\varphi$ = 45º, we can see the walls with $\varphi$ ~ 53/37º which are impermissible in the T phase. These are so called S-walls [42]. Their position is not linked to crystallographic directions, and they are allowed only in low-symmetry phases (all the phases except R and T). By refocusing the microscope, we observed a similar domain structure at different depths inside the crystal plate, confirming that the entire volume of the crystal is in the same phase.

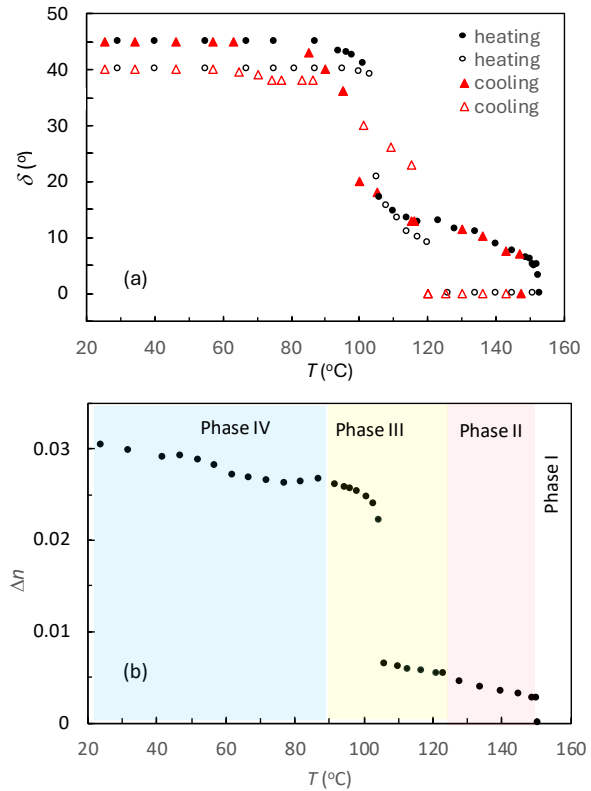

FIG. 8. Temperature dependences of (a) the extinction angles measured on heating and cooling in selected domains of the (001)-oriented PMN-32PT crystal plate, and (b) the birefringence measured on heating in the domains for which the extinction angle is shown by filled circles in panel (a). The temperature ranges of various phases that appear during heating, obtained from the PLM data, are indicated.

Although in most regions of the crystal in Phase II the extinction is not observed, and the domain structure changes with temperature, several domains can be found which exhibit extinction at all temperatures in this phase. In these domains we measured the angles $\delta$ at which the extinction was observed. Figure 8 (a) shows their dependence on temperature. As expected in $M_C$ phase, two values of the extinction angle with $|\delta| < 45°$ are found at each temperature (they are shown by open and filled symbols, respectively), and one of them is $\delta = 0°$. The second extinction angle increases with decreasing temperature. In the $M_C$ phase it is equal to the angle between the axis of optical indicatrix and the [100] axis. The direction of $P_S$ is supposed to be the same (at least approximately) as the direction of one of the optical indicatrix axes. Therefore, the increase of $\delta$ with decreasing temperature means that $P_S$ rotates in the (001) plane away from the [100] direction. This temperature-induced rotation is shown schematically in Fig. 9.

*Contact author: abokov@sfu.ca

†Contact author: zye@sfu.ca

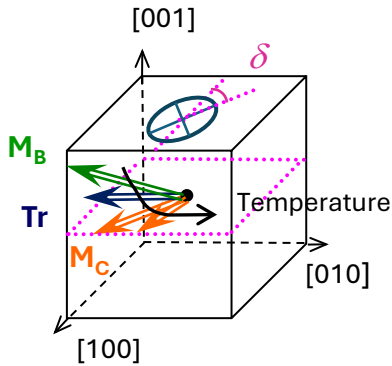

FIG. 9. Schematic of evolution with temperature of the spontaneous polarization direction in Phases IV ($M_B$), III (Tr) and II ($M_C$). The projections of the optical indicatrix and its axes onto the (001) plane are shown.

Figure 8 (b) presents the birefringence ($\Delta n$) as a function of temperature. In general, two values of $\Delta n$ can be expected on a (001)-oriented crystal platelet of the $M_C$ phase, which correspond to the birefringence in domains with extinction angles $\delta = 0°$ and $\delta \neq 0°$, respectively. However, for reliable measurements using a compensator the existence of relatively large domains is required. In the $M_C$ phase of our crystals the domains are typically small, and the accurate measurements are challenging. In practice, we were only able to determine the birefringence of domains with extinction angles $\delta \neq 0°$ [shown by filled circles in Fig. 8 (a)]. As Fig. 8 (b) demonstrates, the jump in $\Delta n$ exists at $T_C$ in accordance with the first order character of the phase transition. In Phase II the measured $\Delta n$ increases linearly with decreasing temperature.

*Contact author: abokov@sfu.ca

†Contact author: zye@sfu.ca

Therefore, the macroscopic symmetry of Phase II is $M_C$, as found previously from optical data in Ref. [12], but not tetragonal as suggested by diffraction data from Refs. [9,13,14] and from Section III.B. Various possible explanations can be considered to reconcile this discrepancy. First, the issue may be associated with the specific domain structure of the crystals we examined. If the domain variants responsible for one of the three $h00$ peaks of the $M_C$ phase are accidentally absent from the crystal volume illuminated by the X-ray beam, only two peaks remain in the diffraction pattern which can be misinterpreted as the T phase. However, we believe that given the very small domain size and complex domain arrangement, such a special domain structure is unlikely to form. Note that according to our and the previously published data, the T symmetry of Phase II is deduced from the diffraction experiments (even in powders where all domain orientations are present), while all PLM experiments suggest the $M_C$ symmetry. Thus, the observed discrepancy is more likely related to the characteristics of the different research methods.

The second scenario to consider is the possible skin effect which would imply that the symmetry of the regions near large faces of the crystal plate is tetragonal, while that of the bulk is monoclinic. The penetration depth of x rays is relatively small, so in XRD experiments a thin layer of ~ 5 μm is probed, which may be thinner than the tetragonal skin. Therefore, the monoclinic phase is not observed. In transmission PLM experiments, which probe both the skin and the bulk, it would be difficult to identify the tetragonal domains in the skin because they exhibit identical extinction directions (at 0/90°) to some of the $M_C$ domains (see Fig. 5), and the domain structure characteristic of tetragonal phase may be barely visible due to the small thickness of the skin. However, such a scenario is also unlikely, because in all known cases of the skin effect in relaxors, the lattice symmetry in the skin is the same or lower than that in the bulk [18,19,46,47]. This behavior could be explained by the surface plane stress which deforms the symmetric unit cell characteristic of the bulk, giving rise to a lower symmetry in the skin layer in relaxors [47]. It is hardly possible for the surface stress to deform a monoclinic unit cell into a tetragonal one. The true reasons for the apparent discrepancy between the diffraction and optical results will be further discussed in the following Section IV.

The transformation to Phase III is accompanied by the emergence of new domain walls in some regions of the crystal, and in the temperature interval of Phase III the new set of walls gradually replaces the old one. Typical PLM images of Phase III are shown in Fig. 6

(g - i). A relatively large area of extinction can be observed in Fig. 6 (i) taken at δ = 20°. Interestingly, at δ = 0/90° and δ = 45° [Fig. 6 (g) and Fig. 6 (h), respectively], no extinction areas can be noticed. This behavior is characteristic of a triclinic phase only (Fig. 5). Similarly, no extinctions at δ = 0/90° and δ = 45° which are forbidden for the Tr phase were observed in the temperature range of Phase III in our examinations of other PMN-32PT crystals. Furthermore, at the same temperature, domains with different extinction angles in the range of 0 < |δ| < 45° can be found [see Fig. 8 (a)], which is possible only in a Tr phase. We conclude, therefore, that Phase III is triclinic. This conclusion is also in agreement with the X-ray experiments which revealed the 400 triplet [Figs. 3(c) and 4)]. The only theoretically possible triclinic polar space group is $P_1$ which should be assigned to Phase III.

The facts that in Phase III the extinction at $\delta$ = 0° disappears and extinction angles continue to increase with decreasing temperature, mean that $P_S$ leaves the (001) plane and continues to move away from the [100] direction, as shown schematically in Fig. 9.

The domain structure of Phase IV differs significantly from those observed in the high-temperature phases. In Fig. 6 (j-n) it is shown for two temperatures, 24°C and 62°C, to demonstrate that it is virtually independent of temperature. Note that many regions remain relatively dark at any $\delta$. The photograph in panel (o) is made without an analyzer to show that these regions are opaque due to the large concentration of domain walls separating small domains, rather than due to extinction. The fact that regions with a large concentration of domain walls inclined to the surface make PMN-PT crystals poorly transparent was demonstrated in Refs. [48,49]. Nevertheless, many transparent regions of the crystal are in extinction at δ = 45° [Fig. 6 (j, m)], which excludes the T, $M_C$ and Tr phases. Besides, the S-walls [Fig. 6 (l)] and the regions that are bright at any $\delta$ are incompatible with the R symmetry. We also observed in Phase IV the domains with temperature-independent extinction at δ = 40° [Fig. 8(a)], which excludes the O phase. Based on these data alone, we cannot distinguish between $M_A$ and $M_B$ phases, for which extinction angles at both δ = 45° and δ ≠ 45°, and S-walls are allowed.

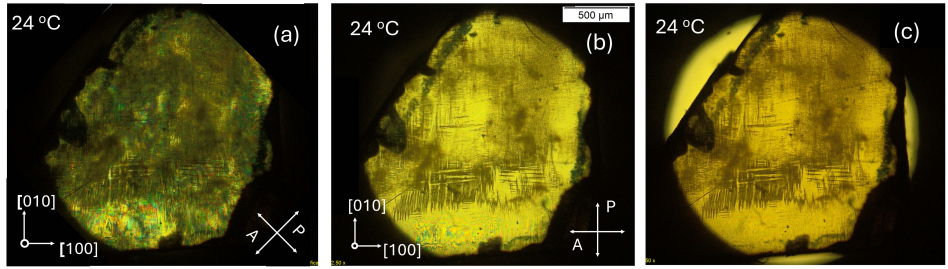

FIG. 10. Polarized light microscopy images of the poled (001)-oriented PMN-32PT crystal at room temperature with crossed polarizers (a, b) and without an analyzer (c). The directions of pseudocubic crystallographic axes and crossed polarizers are indicated.

To further clarify the symmetry of Phase IV, we studied poled crystals. In a fully poled state, the vectors $P_S$ in all domains are directed so that the angle between $P_S$ and the poling field is minimal. Such domain configuration minimizes electrostatic energy. In the case of $M_A$ phase, the poling field along [001] creates the domains with $P_S$ lying in the {110} planes which are perpendicular to the surface of (001) crystal plate. The extinction at δ = 45° should be observed in all regions of the crystal as a result. In the $M_B$ phase this poling creates the domains having $P_S$ in the {110} planes inclined to the surface. In this case, overlapping domains may lead to regions without extinction and in the regions where the extinction is still observed it must be at δ ≠ 45°.

The electroded PMN-32PT crystal was poled by applying a dc electric field with the strength of 6 kV/cm, which is much larger than the coercive field. All (or almost all) the domains were switched by poling into energetically optimal directions, as confirmed by the fact that the measured piezoelectric coefficient $d_{33}$ was about 1700 pC/N, as expected in fully poled PMN-32PT crystals [50]. The PLM images of the poled crystal obtained after removing the electrodes are shown in Fig. 10. No complete extinctions are observed at δ = 0/90° nor at δ = 45°, which excludes the $M_A$ phase and is consistent with the $M_B$ phase. We conclude, therefore, that based on the XRD and PLM results the low-temperature Phase IV in our PMN-32PT crystals is $M_B$. The fact that the crystal looks much darker (closer to extinction) at δ = 45° than at δ = 0/90° confirms that the $P_S$ direction is close to <111>. This is shown in Fig. 9. The temperature-independent extinction at δ = 40° [Fig. 8(a)] means that the temperature-driven rotation of $P_S$ ceases in phase IV and the angle of ~ 5° between $P_S$ and [1-11] direction remains constant.

*Contact author: abokov@sfu.ca

†Contact author: zye@sfu.ca

Based on our XRD and PLM data, we can determine the temperatures of phase transitions. In the temperature range of Tr phase, the 400 triplet and more than one optical extinction should be observed at 0 < |δ| < 45°, while the extinctions at δ = 0/90° and δ = 45° should be absent. This range is found to be approximately from 90°C to 125 °C upon heating and 60°C to 115°C upon cooling [see Figs. 4 and 8(a)]. Interestingly, there are no noticeable changes in the domain structure and no anomalies in the dielectric properties at the transitions between Phases IV, III and II (Figs. 2 and S6). Anomalies in the Δn(T) curves can be observed [Fig. 8 (b)], but they are very small. On the other hand, a large jump in birefringence and a small peak of permittivity is found in Phase III away from the phase transitions (at 105 °C upon heating). These anomalies are caused by two related events observed at that temperature: a sharp change in $P_S$ directions and transformation of the domain structure.

## IV. CONCLUSIONS

Combining the results of our XRD, PLM and dielectric measurements, we conclude that the PMN-32PT crystals exhibit the spontaneous relaxor-to-ferroelectric phase transition. In the paraelectric phase above the Burns temperature, $T_B$, the Curie-Weiss law, Eq. (3) is observed. In the temperature range between $T_B$ and $T_C$ the phenomena characteristic of ergodic relaxor phase are found, including the Vogel-Fulcher law, Eq. (2), the quadratic law, Eq. (1) with γ = 2, the X-ray diffuse scattering and macroscopic cubic symmetry. At $T < T_C$ three ferroelectric phases exist, depending on temperature. The primitive perovskite cells of all phases are shown schematically in Fig. 11. The symmetry of the low-temperature Phase IV is monoclinic $M_B$ (Cm). The same Cm symmetry was previously observed in the PMN-32PT crystals [12], but in ceramics the $M_C$ phase was found, while the $M_B$ phase appeared only at a lower x = 0.3 [10,13]. This means that in ceramics the boundary concentration $x_{B2}$ is slightly lower than in crystals.

The monoclinic Phase IV transforms upon heating to triclinic Phase III. This phase sequence is not normal, since the crystal symmetry generally increases upon heating. However, the transition from monoclinic to triclinic phase upon heating has also been observed using transmission electron microscopy in the MPB region of another lead-containing perovskite system, PZT [51]. In that work curved domain boundaries, unusual for normal ferroelectric phases, were observed and the ferroelectric domains were composed of aggregations of regions about 10 nm in size, having various polarization directions. Curved boundaries also exist in Phases III and IV of our PMN-32PT crystals (Fig. 6), strengthening an analogy with the monoclinic and triclinic phases of PZT. Therefore, it is possible that the macroscopic domains in PMN-32PT are also composed of nanodomains whose symmetry differs from the macroscopic one. This may be one of the reasons why neutron powder diffraction experiments [13], which probe the structure at a smaller scale than PLM revealed the Pm symmetry for Phase III. An alternative possible reason is that the authors of Ref. [13] simply did not perform the Rietveld refinement with a triclinic model, which would provide a better fit than the monoclinic Pm.

The available structural data for Phase II looks contradictory. The published X-ray and neutron powder diffraction studies [9,13,14] revealed the tetragonal P4mm space group, and our XRD data are consistent with this symmetry, showing a splitting of the 400 reflection into two peaks [Fig. 3(d)] as opposed to three peaks generally expected in the $M_C$ (Pm) phase. On the other hand, our PLM results clearly reject the T phase but suggest the $M_C$ (Pm) symmetry, in agreement with the PLM results published in Ref. [12]. To reconcile these observations, we suggest that Phase II is pseudotetragonal, i.e. two lattice parameters, a and b, are equal and the monoclinic angle, β, is close to 90°, as shown in Fig. 11. The small value of β is consistent with the small (~ 5°) deviation of $P_S$ from the [100] direction which we derived from the PLM data, and which implies a small deviation of the monoclinic cell from the tetragonal shape. This unit cell results in the 400 doublet and the small monoclinic distortion that has probably been overlooked in the published refinements of powder diffraction data.

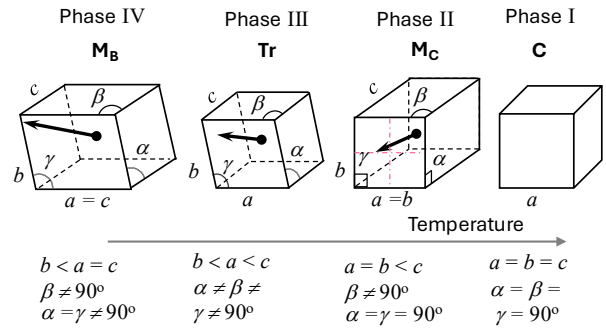

FIG. 11. Schematic of the structural phase transition sequence in the PMN-32PT crystals. Deformations of the perovskite unit cell are exaggerated for clarity. Thick arrow shows the direction of $P_S$.

*Contact author: abokov@sfu.ca

†Contact author: zye@sfu.ca

Therefore, the sequence of phases in the PMN-32PT crystal is $M_B$ ($Cm$) → Tr ($P_1$) → $M_C$ ($Pm$) → C ($Pm\bar{3}m$) upon heating. In all the phases except $M_B$ the Bravais lattice is primitive, thus the parameters of the primitive perovskite cell shown in Fig. 11 coincide with the conventional lattice parameters. In the base-centered $Cm$ lattice the conventional parameters (usually mentioned in structural reports) are related to those shown in Fig. 11 via $c_m = b$, $a_m \approx b_m \approx \sqrt{2}a$, and $\beta_m \approx \alpha$.

Using PLM we observed in the Tr and $M_C$ phases the temperature-driven rotation of spontaneous polarization from a direction close to <111> in the $M_B$ phase to a direction close to <100> near $T_C$, as shown in Figs. 9 and 11. In the $M_B$ phase the polarization rotation is absent. Temperature hysteresis is found for all the phase transitions in PMN-32PT, which suggests that they are all first-order transitions. On the other hand, the triclinic $P_1$ space group of Phase III is a subgroup of both $Pm$ of Phase II and $Cm$ of Phase IV, so that when the temperature changes, a smooth pass between all three phases is theoretically possible (and really observed here) with continuous changes in the lattice parameters and rotation of the spontaneous polarization vector. This explains why at the $M_B$ → Tr and Tr → $M_C$ phase transitions the domain structure does not change, while the permittivity (Fig. 2) and birefringence (Fig. 8) change gradually without significant jumps, characteristic of first-order transitions. In the middle of the temperature range of Tr phase, a jump of birefringence and a sharp dielectric peak are observed, which is due to the sharp rearrangement of the domain structure. This behavior is unusual which may be the reason why in some previous studies the anomalies in properties were interpreted as the result of a phase transition, and the intermediate Tr phase was overlooked.

Future investigations of PMN-PT single crystals, potentially involving advanced X-ray, neutron and electron diffraction as well as spectroscopic techniques, may further elucidate the nature of crystal structure by distinguishing subtle structural differences among phases at different scales, the role of polarization dynamics of nanodomains and their impact on macroscopic properties.

### ACKNOWLEDGMENTS
This work was supported by the U. S. Office of Naval Research (ONR, Grant No. N00014-21-1-2085) and the Natural Sciences and Engineering Research Council of Canada (NSERC, DG, RGPIN-2023-04416). The authors thank Xifa Long and Maryam Bari for their help in crystal growth and sample preparation.

*Contact author: abokov@sfu.ca

†Contact author: zye@sfu.ca

*Contact author: abokov@sfu.ca

†Contact author: zye@sfu.ca

*Contact author: abokov@sfu.ca

†Contact author: zye@sfu.ca

# Supplemental Material

# Temperature-driven polarization rotation and triclinic phase at morphotropic phase boundary of $Pb(Mg_{1/3}Nb_{2/3})O_3 - PbTiO_3$ crystals

Alexei A. Bokov, Haiyan Guo and Zuo-Guang Ye

*Department of Chemistry and 4D LABS, Simon Fraser University, Burnaby, BC, V5A 1S6, Canada*

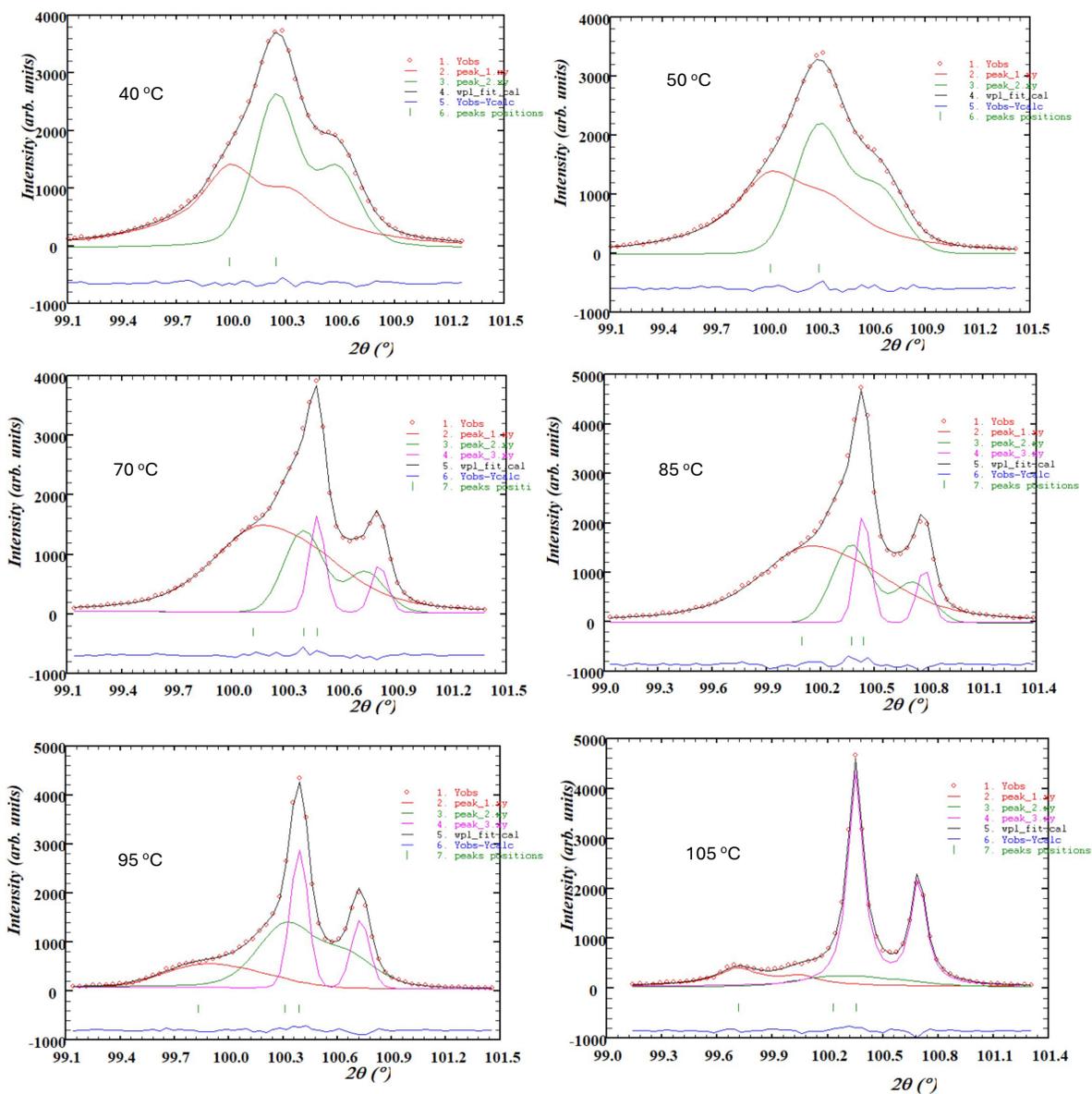

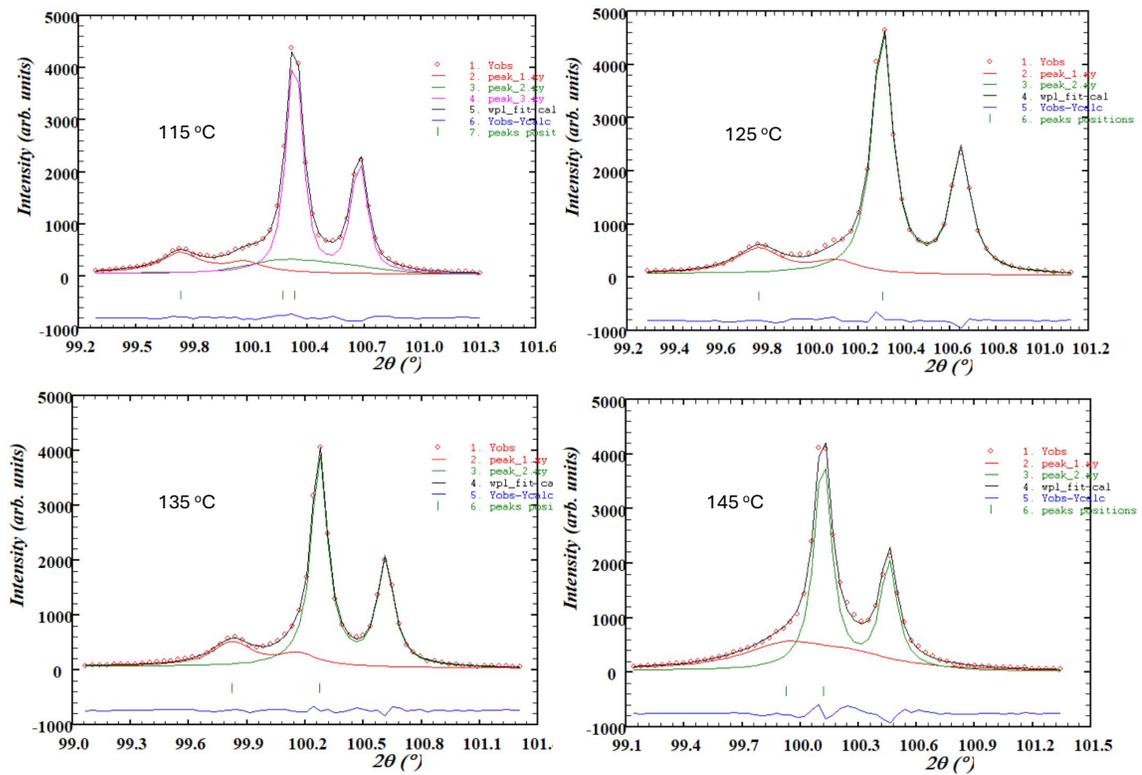

FIG. S1. The profiles of the 400 Bragg peak obtained from a main facet of PMN-32PT crystal plate at different temperatures during slow cooling. The K$\alpha_2$ component is not removed. The black line is the least-squares fit to the data points shown by circles, and the colored lines are fitted pseudo-Voigt subpeaks.

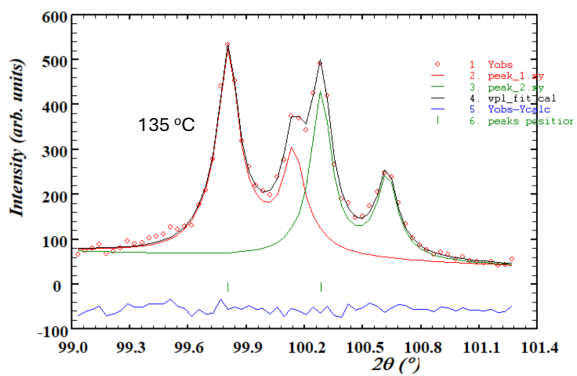

FIG. S2. The profile of the 400 Bragg peak obtained from a main facet of PMN-32PT crystal plate in Phase II at 135 °C during slow heating. The K$\alpha_2$ component is not removed. The black line is the least-squares fit to the data points shown by circles and the colored lines are fitted pseudo-Voigt subpeaks.

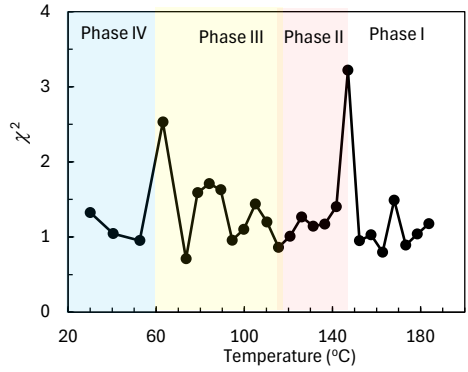

FIG. S3. Temperature dependence of the parameter $\chi^2$, which characterizes goodness of fit to tree (for Phase III) or two (for other phases) pseudo-Voigt subpeaks for the 400 Bragg reflection.

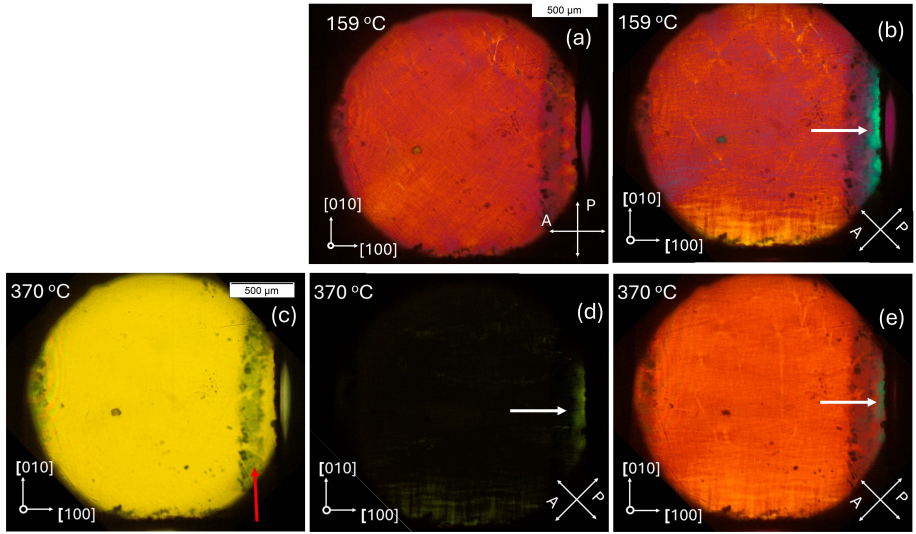

FIG. S4. Polarized light microscopy images of the same (001)-oriented PMN-32PT crystal plate as shown in Fig. 6 of the main text obtained during heating in the paraelectric phase close to the Curie point at 159 ºC (a, b) and at 370 ºC (c-e). The directions of crystallographic axes and crossed polarizers are indicated. In photographs (a, b, e) the first order red plate is superimposed. Photograph (c) made without polarizers demonstrates the transparency of the crystal. The vertical partially transparent stripe, approximately parallel to the crystal's edge and indicated by the red arrow, is the remnant of the gold electrode. Photograph (d), taken with polarizes angle $\delta = 45°$, demonstrates that at a temperature as high as 370 ºC the region near the crystal edge (marked with the white arrow) remains birefringent (bright). This region is better seen in Photograph (e), where it is colored greenish blue. At lower temperatures it is wider, as in Photograph (b), taken at 159 ºC with polarizes at the angle $\delta = 45°$. When $\delta = 0/90°$ [Photograph (a)] the entire crystal is in complete extinction (it has a magenta color, the same as the non-birefringent area outside the crystal plate). Therefore, in the paraelectric phase the near-surface birefringent layer exists under the small (100) facet of the crystal plate. The thickness of this layer is about 50 -100 μm and the extinction is compatible with the tetragonal symmetry. The retardation of ~ 100 nm is found in the surface layer using the Michel-Levy interference color chart, which corresponds to the birefringence $\Delta n = 0.0002$.

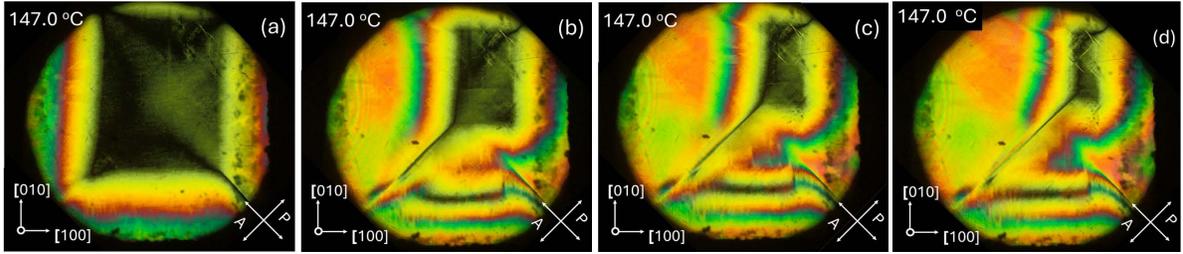

FIG. S5. Isothermal motion of the phase front at the Curie point observed in polarizing microscope on the (001)-oriented PMN-32PT crystal plate. The temperature was stabilized at 147.0 °C after slow (0.5 K/min) cooling from the paraelectric phase, and the photographs (a-d) were taken at 4 min intervals. The dark region in the center is in the paraelectric phase, while the multicolored birefringent regions are in the ferroelectric phase.

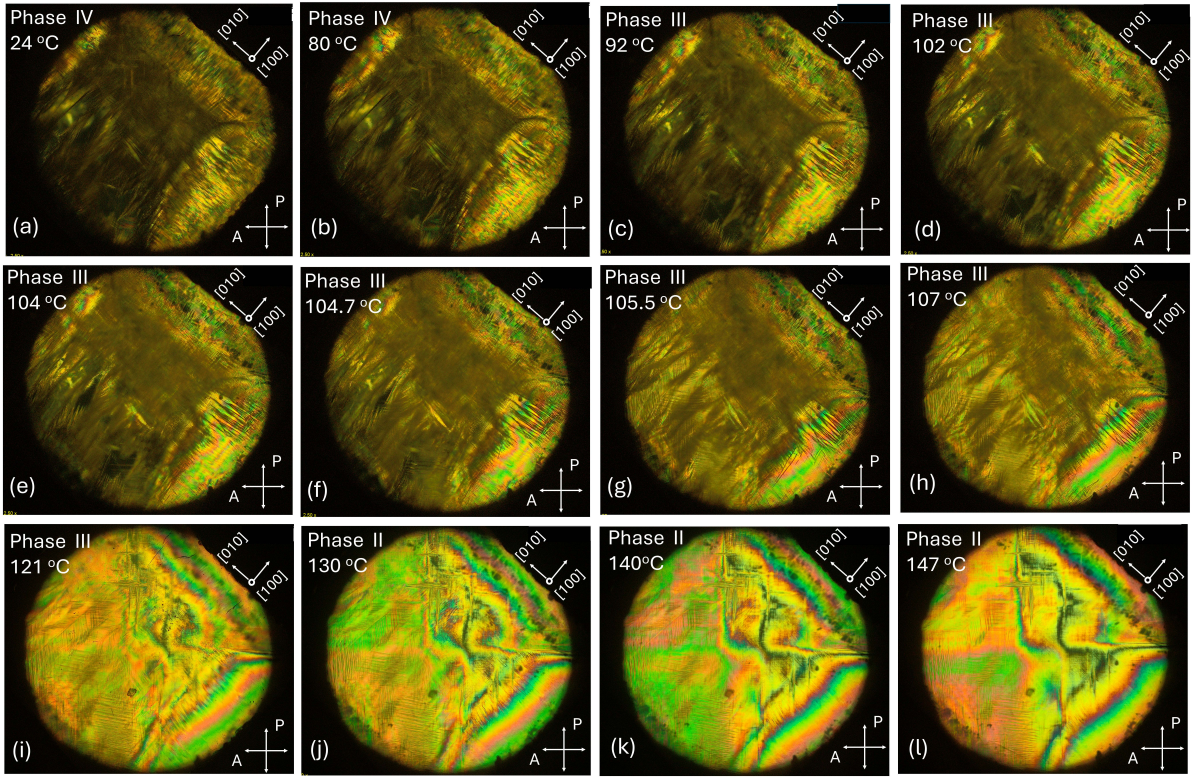

FIG. S6. Evolution of the domain structure in the (001)-oriented PMN-32PT crystal plate observed under a polarizing microscope during heating. The temperature, directions of pseudocubic crystallographic axes and crossed polarizers are indicated. The platelet thickness is 540 μm. The configuration of domain walls does not change noticeably at phase transitions (90°C and 125 °C) but changes sharply around 105 °C. The color changes seen in panels (i-l) are due to the variation of birefringence rather than domain structure.